\documentclass[twocolumn,11pt]{article}
\usepackage[margin=0.75in]{geometry}
\usepackage{amsmath, amssymb}
\usepackage{graphicx}

\usepackage{caption}
\usepackage{authblk} 
\usepackage{physics}
\usepackage{abstract}
\usepackage{siunitx}
\usepackage{bm}
\usepackage{tikz}

\usepackage{xcolor}
\RequirePackage[colorinlistoftodos,prependcaption,textsize=scriptsize,backgroundcolor=orange!20]{todonotes}

\definecolor{mygreen}{rgb}{0.0,0.5,0.0}
\definecolor{myblue}{rgb}{0.0,0.0,0.65}
\definecolor{myred}{rgb}{0.72,0.0,0.}
\definecolor{mycyan}{rgb}{0.0,0.6,0.6}
\definecolor{mylila}{rgb}{0.6,0.,0.6}
\definecolor{mygray}{rgb}{0.37,0.37,0.37}

\usepackage[compact]{titlesec}
\titlespacing{\section}{0pt}{10pt}{4pt}
\titlespacing{\subsection}{0pt}{8pt}{3pt}

\usepackage{soul}\usepackage[normalem]{ulem}

\usepackage[
    backend=biber,
    style=phys,     
]{biblatex}

\usepackage{hyperref}

\title{\bfseries Kinematic Fitting of  electromagnetic calorimeter data \\ - an improved method}
\author[1*]{Nicolas Kolanus}
\author[1]{Jonas Kohlen}
\author[1]{Ulrike Thoma}
\affil[1]{Helmholtz-Institut für Strahlen- und Kernphysik and Cluster of Excellence "Color meets Flavor", Universität Bonn, Germany}
\affil[*]{Corresponding author: \texttt{kolanus@hiskp.uni-bonn.de}}

\date{}
\begin{document}

\twocolumn[
\maketitle
\begin{onecolabstract}
\noindent Kinematic fitting is widely used in particle physics experiments as a powerful tool to improve  experimental resolutions, to suppress background and to provide selection criteria for the identification of specific reactions. 
Kinematic fitting methods, however, typically assume that the fitted quantities follow Gaussian distributions; an assumption which does not always hold. This is particularly true for energy measurements from electromagnetic calorimeters. 
This issue can largely be overcome by performing a transformation of the non-Gaussian variables into variables which follow a Gaussian distribution. The new adapted kinematic fitting procedure allows to better account for the non-Gaussian nature of the measured calorimeter energies, at the same time improving the accuracy and robustness of the kinematic fit. 
This leads to improved pull and confidence-level distributions of the kinematic fit as well as improved invariant mass distributions and signal-to-background ratios in the final event sample.   
\end{onecolabstract}
\vspace{1em}
]

\section{Introduction}
In particle physics experiments, kinematic quantities of final state-particles are measured as part of reconstructing the reaction of interest. Of course, these measurements are subject to experimental uncertainties which limit the resolution of derived quantities such as the invariant masses of intermediate particles. In addition, background may dilute the process under investigation. 
To improve resolution, suppress background, and  reconstruct 
undetected particles, kinematic fitting procedures are commonly applied. These  enforce physical constraints such as the conservation of energy and momentum.  
During kinematic fitting, the measured values are adjusted within their uncertainties to find the most likely configuration that satisfies known constraints. Beyond improving the precision of reconstructed quantities, this technique also provides a statistical measure of the fit quality via the $\chi^2$ value. This provides a measure of how likely an event stems from the specific reaction of interest. As a mathematical problem, this procedure is well understood and even outlined in textbooks \cite{brandtDataAnalysisStatistical2014}. 
However, a limitation of classical kinematic fitting is its reliance on the assumption that the uncertainties of all fitted quantities follow Gaussian distributions. In practice, this assumption is often violated. Notably, energy measurements performed by electromagnetic calorimeters (EMCs) typically exhibit an asymmetric non-Gaussian shape (see e.g.~\cite{IKEDA2000}).\\

\noindent
In this article, we discuss a newly-developed method which transforms the measured energies such that the distributions of the transformed variable closely approximate a Gaussian distribution. The aim of the new approach is to make kinematic fitting more robust and statistically consistent in scenarios where non-Gaussian input uncertainties would otherwise introduce significant distortions.\\

\noindent
The paper is organized as follows: In section~\ref{ch:theory} a brief introduction to kinematic fitting is given. Section~\ref{ch:statistics} discusses the statistical concepts and terminology used in section~\ref{ch:transform}, where the transformation method itself is introduced.
Finally, section~\ref{ch:Test} tests the new method on simulated data of the Crystal Barrel calorimeter of the CBELSA/TAPS experiment.

\section{Kinematic fitting}
\label{ch:theory}
\noindent
The goal of kinematic fitting is to constrain the events measured in an  experiment to a specific reaction hypothesis, typically via $\chi^2$ minimization. The reaction hypothesis is defined in terms of physical constraints such as energy and momentum conservation and/or the masses of the decaying particles (e.g. $\pi^0 \to \gamma\gamma$ or $\eta \to \gamma\gamma$). 
This procedure can be reduced to a minimization problem where the measured quantities $x_{0,i}$ are varied within their respective uncertainties and the resulting fit values of $x_i$ with $1\leq i \leq n$  should then minimize $\chi^2$ while simultaneously fulfilling $m$ constraint equations $f_j(\vec{x}) = 0$ where $1\leq j \leq m$. Given the covariance matrix $G^{-1}_{ik}$ of the measured quantities, the $\chi^2$ can be written as:
\begin{equation}
    \chi^2 =(x-x_0)_iG_{ik}(x-x_0)_k = \delta_{i} G_{ik} \delta_{k}
    \label{eq:chi2}
\end{equation}
where $\delta_{i}$ is the difference between the fitted values $x_i$ and the measured values $x_{0,i}$. If $f_j$ are linear functions of $x_i$, the conditions can always be expressed as:
\begin{equation}
    F\vec{\delta}-\vec{d}=0
    \label{eq:condition}
\end{equation}
where $F$ is a $m\times n$ matrix of rank $m$ and $\vec{d}$ is a vector of dimension $m$. We only consider cases where $m<n$ as then Eq. (\ref{eq:condition}) does not have a unique solution, but rather an $n-m$ dimensional space of solutions. Using a similar notation, the $\chi^2$ can be rewritten as:
\begin{equation}
    \chi^2 = \delta_iG_{ik}\delta_k = (A\vec{\delta})^2 \, .
    \label{eq:Ab}
\end{equation}
Thus, the fit can be reduced to determining the $\vec{\delta}$ in the subspace of solutions of $F\vec{\delta}-\vec{d}=0$ which minimizes $(A\vec{\delta})^2$. This is a well known linear algebra problem, the solution of which can be directly calculated. For this, the method of Lagrange multipliers is often utilized, but for all results shown in this paper, a method based on orthogonal transformations has been used, as this has been found to have better numerical stability. A description of the algorithm can be found in the \hyperref[ch:Appendix]{Appendix}.

\subsection{Linearization and Iteration}
\noindent
In the case of linear constraints, the solution of the minimization problem can be computed directly. However, the particle physics constraints used in kinematic fitting are in general not linear. Thus, the solution cannot be calculated directly using the discussed minimization methods. \\   

\noindent
But even in the case of non-linear constraints, a solution can be obtained by approximating the non-linear problem as a linear one. The solution is then approached iteratively by using the result of each fit as a starting point for the  next iteration. The iterative procedure is continued until the $\chi^2$ does not change significantly between iterations, or it is terminated after a fixed number of steps has been reached due to failure to converge.\\
Performing an expansion around the measured point $x_{0}$ for the first iteration, the matrix $F$ and the vector $\vec{d}$ in Eq.~(\ref{eq:condition}) are given to first order by:
\begin{equation}
    F_{ij}= {\left. \frac{\partial f_i}{\partial x_j} \right |_{\vec{x}=\vec{x}_{0}}} \quad \quad d= -f(\vec{x}_{0}) \, .
    \label{eq:derivatives}
\end{equation}
For all but the first iteration, the equations are slightly adjusted as one has to differentiate between the start values of the fit procedure and the measured value $x_{0,i}$. The start value of the iteration, $x_{s,i}$, is always the fit result of the previous iteration\footnote{For the first iteration $\vec{x}_s=\vec{x}_0$.}. 
Therefore, the expansion of the condition functions $f_{j}$ is performed around $x_{s,i}$, such that Eq.~(\ref{eq:derivatives}) is modified to:
\begin{equation}
    F_{ij} = {\left . \frac{\partial f_i}{\partial x_j} \right |_{\vec{x}=\vec{x}_{s}}} \quad \quad d= -f(\vec{x}_s) \, .
    \label{eq:derivatives_updated}
\end{equation}
In addition, $\vec{\delta}$ (Eq.~(\ref{eq:condition})) is redefined as:
\begin{equation}
    \label{eq:redef_delta}
    \vec{\delta} = \vec{x}-\vec{x}_{s} \, .
\end{equation}
At each iteration, the $\chi^2$  has to be calculated with respect to the measured values $x_{0,i}$, leading to modifications of Eq.~(\ref{eq:chi2}) and Eq.~(\ref{eq:Ab}):
\begin{equation}
    \label{eq:modified_chi2}
    \chi^2=(A\vec{\delta}-\vec{b})^2
\end{equation}
with $\vec{b}=A(\vec{x}_{0}-\vec{x_s})$.

\subsection{Goodness-of-Fit}
\label{GoF}
\noindent
By construction, the $\chi^{2}$ minimization procedure assumes that the measured quantities are distributed symmetrically around their means. 
If, in addition, the distributions are Gaussian and the number of degrees of freedom (d.o.f.) is known, the probability value $p$ can be computed from the $\chi^{2}$. 
We refer to the probability value $p$ as the confidence level (CL), which is given by 
\begin{equation}
    \text{CL} = 1 - \int_0^{\chi^2}f_\text{d.o.f.}(\tilde{\chi}^{2} )\mathrm{d}\tilde{\chi}^2 \, .
    \label{eq:CL}
\end{equation}
Assuming that the fitted hypothesis is correct, the CL gives the probability of sampling an event with the given $\chi^2 $ or larger. Note that the CL does not give the probability that the hypothesis is correct. In the ideal case, as discussed here, the CL is a useful selection criterion, as it is always known how much signal is lost.  For example, a cut rejecting every event with confidence smaller than $0.1$ rejects 10\% of signal events.\\

\noindent
Another statistical quantity that can be used to assess the quality of the fit is the pull. This is the normalized distance of the fitted value to the measured value:
\begin{equation}
    \mathrm{pull_i} = \frac{x_i-x_{0,i}}{\sqrt{\sigma_{x_{0,i}}^2-\sigma_{x_i}^2}}
    \label{eq:pull}
\end{equation}
where $\sigma_{x_0}$ is the uncertainty of the measured quantity and $\sigma_x$ is the uncertainty of the fit result of the respective quantity. The pull distribution is expected to follow a normal distribution (centered around zero with a standard deviation of one). All of these statements heavily rely on the fact that the distributions of the measured variables are Gaussian-distributed and that their errors are well known - otherwise the interpretation of the confidence level will not hold. The confidence level distribution of the signal events will not be flat, and the pull distributions will not follow a normal distribution. \\

\noindent
In energy measurements of electromagnetic calorimeters this assumption is not fulfilled. The  measurements do not follow Gaussian distributions but rather feature long tails to low energies \cite{IKEDA2000}, which are caused by energy losses. The distribution can be described by the Novosibirsk function, for which the following parametrization has been used:
\begin{equation}
    f(x)=\exp\left(-\frac{1}{2\eta^2}\ln^2\left(1+\frac{(x-\mu)}{\sigma}\cdot\Lambda\right)-\eta^2\right)
    \label{eq:Novosibirsk}
\end{equation}
where $\Lambda$ is $\sinh(\eta\sqrt{\ln(4)})/\sqrt{\ln(4)}$.
The parameter $\mu$ gives the position of the most probable value (MPV) of the distribution, $\sigma$ controls the width, and $\eta$ gives the severity of the tail of the distribution. In the limit $\eta \to 0$, a Gaussian is obtained, and the sign in front of $\eta$ determines the direction of the tail. As an example, we consider the reconstructed energies in the Crystal Barrel calorimeter. The reconstructed energy distribution extends towards smaller values, corresponding to a negative tail parameter.\\

\begin{figure}[h!]
    \centering
\includegraphics[width=\columnwidth]{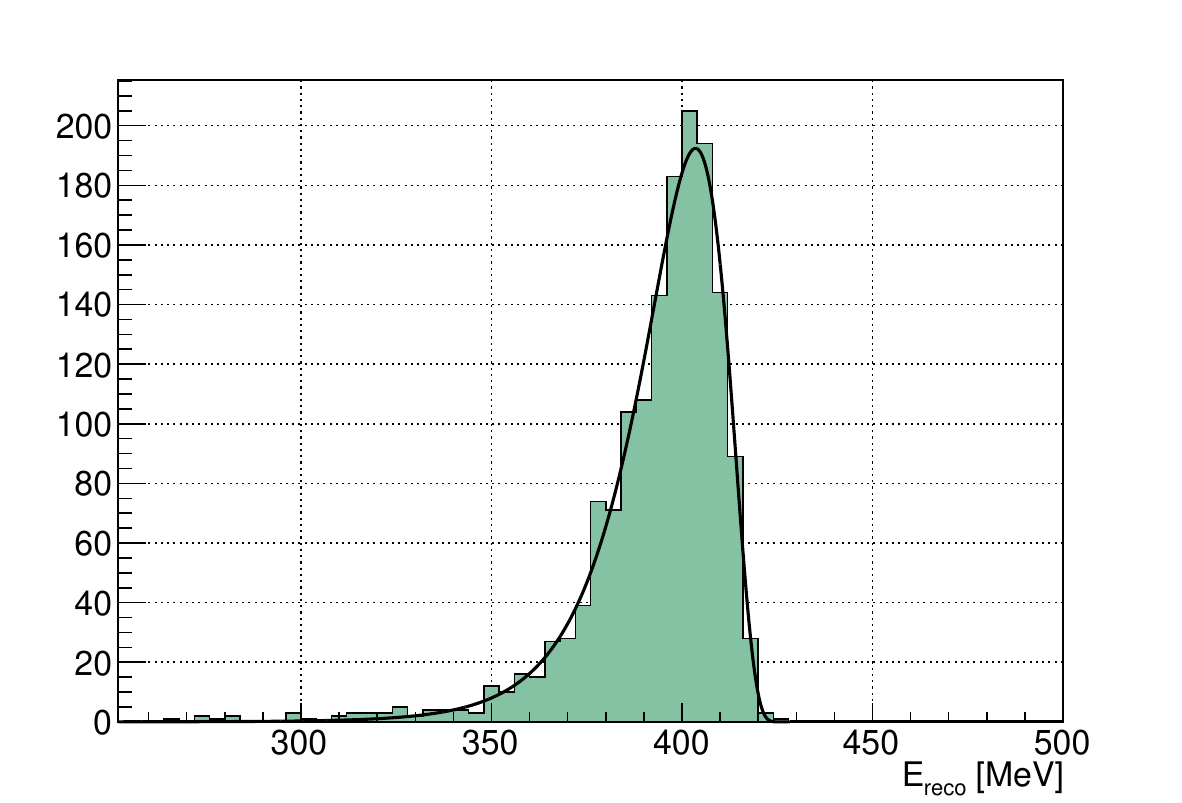}
    \caption{Reconstructed energies $E_{\mathrm{reco}}$ for photons simulated with an energy of $\SI{400}{\mega \electronvolt}$. 
    The black line shows a fit with the Novosibirsk function introduced in Eq.~(\ref{eq:Novosibirsk}).}
    \label{fig:energypdf}
\end{figure}

\noindent
Fig.~\ref{fig:energypdf} shows a fit of Eq.~\ref{eq:Novosibirsk} to reconstructed simulated events  in the Crystal Barrel calorimeter. For the shown spectrum, an energy correction has been applied in the reconstruction such that the MPV of the distribution lies on the true energy. The uncorrected energy values are all smaller than the true energy due to energy losses.\\

\begin{figure}[h]
    \centering
    \includegraphics[width=\columnwidth]{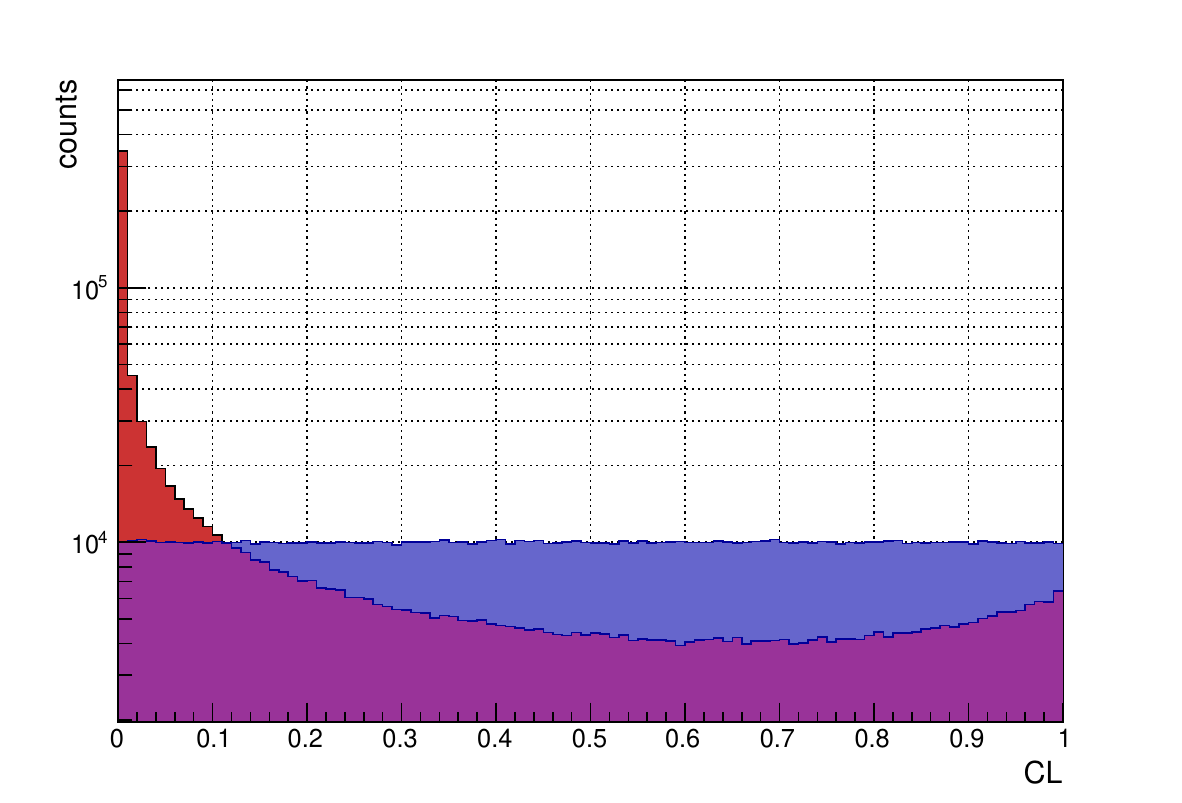}
    \caption{Toy model confidence level computed using values sampled from Gaussian distributions (blue) and Novosibirsk (see Eq.~(\ref{eq:Novosibirsk})) distributions (red). For the Novosibirsk case, the parameter $\sigma$ was used as the error for the calculation of the $\chi^2. $}\label{fig:cl_toymodel}
\end{figure}

\noindent
To illustrate the challenges of asymmetric and non-Gaussian distributions, we calculate the confidence level for two toy models of energy measurements in a calorimeter (see Fig.~\ref{fig:cl_toymodel}). For the two models, values were sampled from either Gaussian distributions with known standard deviation $\sigma$ or from Novosibirsk distributions with known width parameter $\sigma$ and a tail parameter of $\eta=-0.5$, which is a realistic value for the spectra found in electromagnetic calorimeters. From the sampled points, the $\chi^2$ relative to the most probable values of the distributions has been  calculated, which can be translated into the confidence level as shown in Eq.~(\ref{eq:CL}). As can be seen, the observed distribution for the Gaussian case is exactly flat, as we would expect from the definition of the confidence level. However, for the Novosibirsk case,  the confidence level distribution is strongly concave and has a large excess at very low confidence levels. 
This leads to an undesirable situation where many signal events are identified as background events, and there is no simple mapping to the true fraction of signal events that is removed with a given confidence-level cut. Thus, the power of the confidence level cut is seriously reduced.
\section{Statistical Model}
\label{ch:statistics}
\noindent
In the following section, we clarify the distinction between the probability given by a probability density function (PDF) and the likelihood given by a likelihood distribution. The PDF describes the probability of measuring a certain value $X_0$ given the true physical value $X_{\text{true}}$. The likelihood distribution can be regarded as the opposite; given a measured value of $X_0$, the likelihood distribution gives the chance that the corresponding true value is a certain $X_{\text{true}}$. As the PDF is given relative to the true value, it can only be known exactly if the true value is also known. This is, of course, not the case in real experiments, where one only has the measured value to work with.\\ 

\noindent
When performing a fit using the measured values (for example, using the least-squares method), one needs to estimate the error of the measurement, which would be given by the width of the PDF. As the PDF from which the measured value was sampled is generally not known, one needs to estimate this width using the measured value \footnote{This is similar to what is known as Neyman-$\chi^2$, where the variances are given by the measured values.}.
\begin{figure}[h]
    \centering
    \includegraphics[width=\columnwidth]{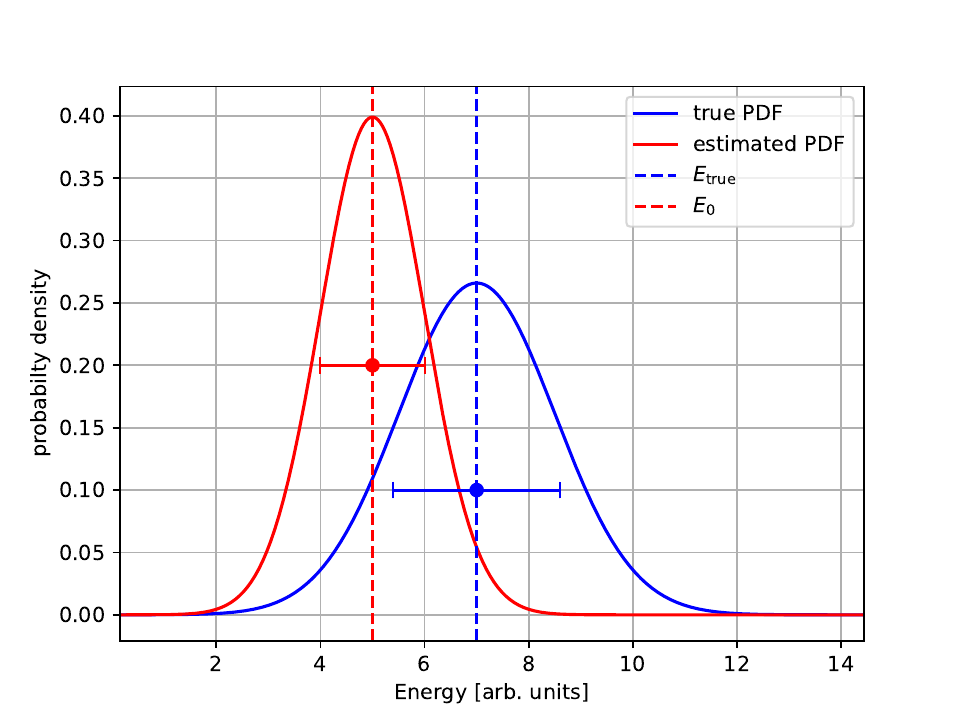}
    \caption{Estimate of the underlying true PDF using a single measured energy $E_0$. The true PDF from which the measurement is sampled is shown in blue and is centered around the true value $E_{\text{true}}$. In this example the precision of the energy measurement depends strongly on the energy. This  leads to the fact that the estimated PDF (determined by assuming a Gaussian with $\sigma=A\sqrt{E_0}$ and $\mu=E_0$), shown in red, is significantly narrower. The $1\sigma$ errors of both the estimated, as well as the true PDF are indicated with the bars.}
    \label{fig:pdf_estimate}
\end{figure}

\noindent
We illustrate this using the example of a calorimeter measurement of the energy of a photon, $E_{\text{true}}$. The detector resolution is often described by a Gaussian PDF with mean $E_{\text{true}}$ and width $\sigma$. For $\sigma$ we assume, in our example, the simplified relation $\sigma=A\sqrt{E_{\text{true}}}$, where $A$ is a known constant specific to that detector. The measurement process can now be described by sampling a random number from this PDF, which we call $E_{0}$. The error of this measurement is then determined using the relation above.\\

\noindent
This situation is shown in Fig.~\ref{fig:pdf_estimate}, where a value $E_0$ between $1\sigma$ and $2\sigma$ below the true value was measured. When using the sampled value as an estimator for $E_{\text{true}}$, the error (for this measurement) is underestimated via $\sigma=A\sqrt{E_0}$, and the MPV of the PDF is assumed to align with $E_0$. Note that for reasonably small relative errors, the differences between the estimated and true errors will usually be smaller than shown in Fig.~\ref{fig:pdf_estimate}, as the effect is exaggerated for the purpose of visualization.\\

\begin{figure}[h]
    \centering
    \includegraphics[width=\columnwidth]{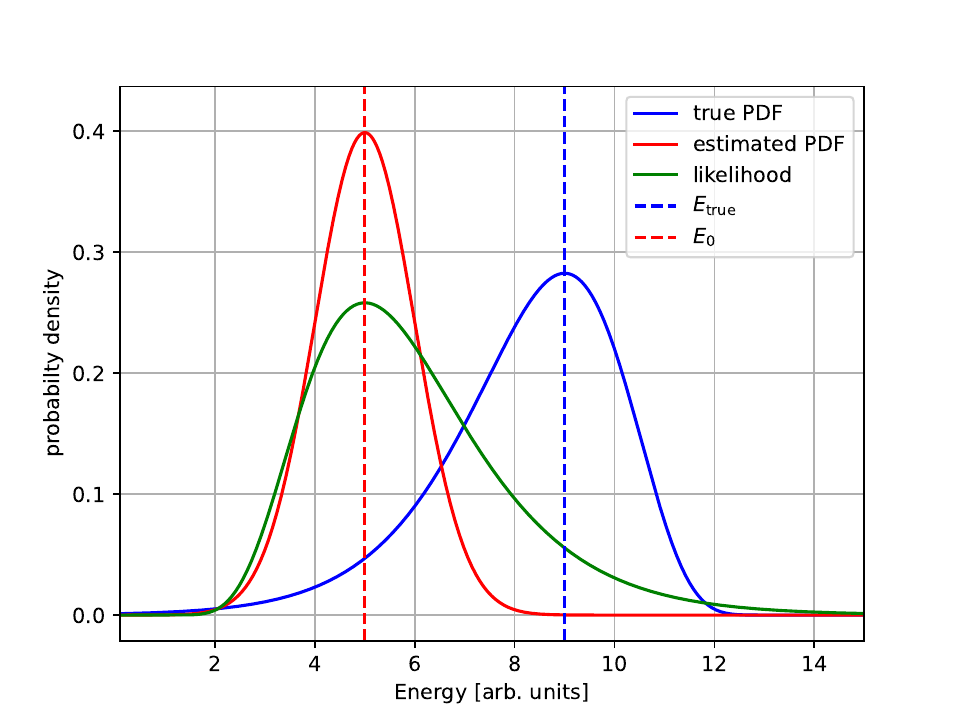}
    \caption{Gaussian estimate (red) of an asymmetric PDF (blue) using the measured energy $E_0$. Additionally shown is the likelihood (green) which gives the chance of the true value being a certain energy given the measured value $E_0$.}
    \label{fig:pdf_estimate_novo_likelihood}
\end{figure}

\noindent
In the case of measurements with real EMCs, however, one will encounter similar problems even for small relative errors. As discussed in Sect.~\ref{GoF}, energy measurements in EMCs do not follow a Gaussian PDF but instead feature a long tail extending to low energies (see Fig.~\ref{fig:pdf_estimate_novo_likelihood}). While here the measured value $E_{0}$ is still quite likely to occur based on the actual PDF, a Gaussian approximation provides a very poor description of the true shape of the PDF, and the true energy value would seem extremely unlikely given the measured value and the estimated error. Replacing the Gaussian with a Novosibirsk would be even worse, as its tail would be pointing away from the true value $E_{\text{true}}$. \\

\noindent
What should be determined instead is the likelihood distribution of true values based on the measured value $E_0$. This is indicated in green in Fig.~\ref{fig:pdf_estimate_novo_likelihood}. Note that although the likelihood looks very similar to the true PDF with flipped tail parameter, the absolute value of the tail parameter $\eta$ and the width parameter $\sigma$ are generally not the same. \\

\noindent
Fig.~\ref{fig:pdf_estimate_novo_likelihood} shows that when performing a kinematic fit based on the measured values, the likelihood distribution represents a much better estimate of the Goodness-of-Fit since it models the statistical nature of the measurement process more accurately. 

\section{Transformation Method}
\label{ch:transform}
\noindent
Given the asymmetric shape of the measured energy distributions in electromagnetic calorimeters, the kinematic fit needs to be adapted to accommodate the mismatch between the assumed Gaussian shapes in the fit and the realities of experimental measurements. This problem can be largely overcome by performing the fit not on the physical energy variable $E$ but on a transformed variable $E'$. \\

\noindent
As outlined in chapter~\ref{ch:statistics}, the PDFs are generally not accessible because the true energy value $E_{\text{true}}$ is not known. In addition, the estimation of the PDFs through the measured parameters is not very sensible due to the highly asymmetric distributions. This means the transformation should not be defined with respect to the estimated PDF but rather on the likelihood distribution, such that the likelihood distribution of the transformed variable becomes Gaussian.\\

\noindent
To achieve this, the 
transformation function $t_{\eta}(E)$ is defined such  that it maps the n-th quantile of a Novosibirsk likelihood distribution with tail parameter $\eta$ onto the n-th quantile of a Gaussian distribution. The details of this function are discussed in section~\ref{sec:function}.\\

\noindent
Furthermore, because photons of different energies exhibit different likelihood distributions, and electromagnetic calorimeters are generally not completely homogeneous, these will also depend on the direction of the photons. Therefore, these dependencies need to be considered carefully, and each photon must be transformed with an individual function. This is achieved by extracting the parameters of the likelihood functions for all measured photons based on simulation, which will be described in section~\ref{sec:likelihood}.\\
 
\subsection{The Transformation Function}
\label{sec:function}
\begin{figure}[h]
    \centering
    \includegraphics[width=\columnwidth]{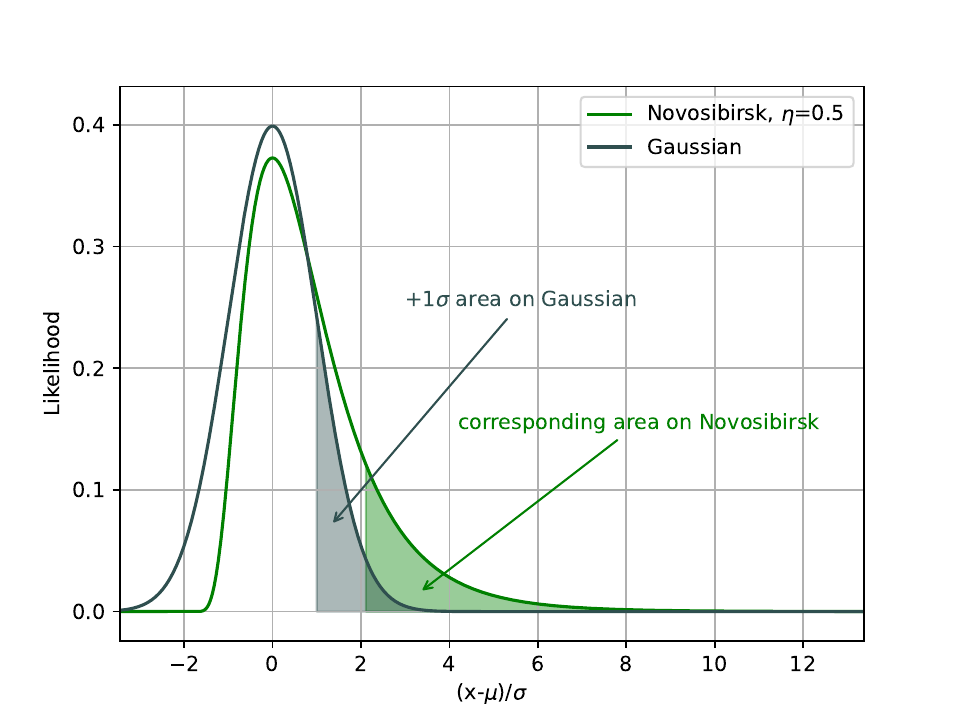}
    \caption{Comparison between a Novosibirsk function with tail parameter $\eta=0.5$ and a Gaussian. The area above $+1\sigma$ on the Gaussian represents the highest $16$\% of sampled events, the corresponding area on the Novosibirsk is also shown, starting $x=2.05\sigma$.}
    \label{fig:novovsgaus}
\end{figure}

\noindent
The desired transformation function needs to transform the likelihood distribution (described by a Novosibirsk function) to a Gaussian. This is achieved by mapping the positions of the same quantiles of both functions onto each other. Fig.~\ref{fig:novovsgaus} shows the area which corresponds to the $16$\% of photons sampled with the highest (energy) values from the Novosibirsk function. The same area for a normal distribution corresponds to the area above   approximately $+1\sigma$. This means that the transformation $t$ should map the $\approx+2.05\sigma$  distance from the MPV in the energy variable (Novosibirsk, corresponding to $16$\%) to the $+1\sigma$ distance from the MPV in the Gaussian variable.

\noindent
To determine this relation for every point, one must compute the cumulative distribution functions $F(x)$ of a normalized Gaussian and a normalized Novosibirsk function ($\mu=0$ and $\sigma=1$) with their respective inverse functions $F^{-1}(x)$, where $x$ is the normalized variable $\frac{E-E_0}{\sigma}$. Using these, the transformation is given by:
\begin{equation}
    t(E) = \sigma_G F_{\text{Gauss}}^{-1}\left(F_{\text{Novo},\eta}\left(\frac{E-E_0}{\sigma_{N}}\right)\right) + E_{0} \, .
    \label{eq:transformation}
\end{equation}
The inverse transformation follows directly as:
\begin{equation}
    t^{-1}(E') = \sigma_N F_{\text{Novo},\eta}^{-1}\left(F_{\text{Gauss}}\left( \frac{E'-E_{0}}{\sigma_G}\right)\right)+E_{0}
    \label{eq:inverse_trafo}
\end{equation}
where $E_0$ is the initial measured energy and $\sigma_N$ and $\eta$ are the parameters of the likelihood distribution fitted by a Novosibirsk function. The width $\sigma_G$ and mean $\mu_G$ of the Gaussian can be freely chosen, as one can scale and shift the transformed variable $E'$ as desired. For reasons of symmetry, we set $\sigma_N=\sigma_G=\sigma$ and define the transformation such that the likelihood distribution in the transformed energy variable is a Gaussian with width $\sigma$ and mean $\mu_G=E_0$.\\

\noindent
For the Gaussian, the cumulative distribution function is well known and is related to the error function $\mathrm{erf}(x)$. For the Novosibirsk function, the cumulative distribution function $F_{\text{Novo,$\eta$}}(x)$, and by extension $t(E)$, must be computed numerically by performing the integration over the Novosibirsk function. To save computational resources at runtime of the kinematic fit, this is done in advance. For different values of $\eta$, the function $t(E)$ is calculated at many points in $x$,  between which splines are used for interpolation. The inverse function $t^{-1}(E')$ can then easily be obtained by exchanging the $x-$ and $y-$axes for the calculated points and performing a similar interpolation.\\

\noindent
When performing the numerical calculation, one must decide on a sensible range over which to perform the calculation, as only a finite number of points can reasonably be calculated. To extend beyond the chosen range, the function needs to be extrapolated. However, by choosing a range that encompasses all reasonably expected events, issues with extrapolation can be minimized. \\

\noindent
When choosing the range over which to compute $t(E)$, we need to account for the fact that the Novosibirsk function becomes and remains zero on the side opposite its tail. That means that, in contrast to the Gaussian, which is non-zero on the entire real number axis, the Novosibirsk function with a positive tail parameter $\eta$ is only non-zero on the interval $(\xi_0,\infty)$, where $\xi_0$ is the point where the function becomes zero. As the transformation $t$ maps quantiles to quantiles, the $(\xi_0,\infty)$ interval is mapped onto the complete real number axis.\\

\noindent
In choosing the computation range, we restrict the quantiles to be between $1\times10^{-6}$ and $1-1\times10^{-6}$, which corresponds to a bit less than $\pm5\sigma$ in the Gaussian variable. This is reasonable as a negligible number of signal events lie outside this range. Beyond this range, the function is extrapolated linearly using the slope between the last two computed points. The following considerations were instrumental in choosing this approach:\\
\begin{itemize}
\item[(i)] Computing an even larger range would be numerically more difficult, as even smaller quantiles would require more careful consideration of integration precision.\\
\item[(ii)] For the quantile mapping to be exact, $t(E)$ has to go to $-\infty$ when approaching $\xi_0$ from above. This means that the fit can never produce results that lie below $\xi_0$, as these values can never be reached even when going to $-\infty$ in the transformed variable. Such a rigid criterion is not necessarily desirable, as the Novosibirsk fit function is still just a model that reduces the full complexity of the energy depositions to a function with a small number of parameters. So using the extrapolation after a certain point allows these values, in principle, to still be reached.
\end{itemize}

\begin{figure}[h]
    \centering
    \includegraphics[width=\columnwidth]{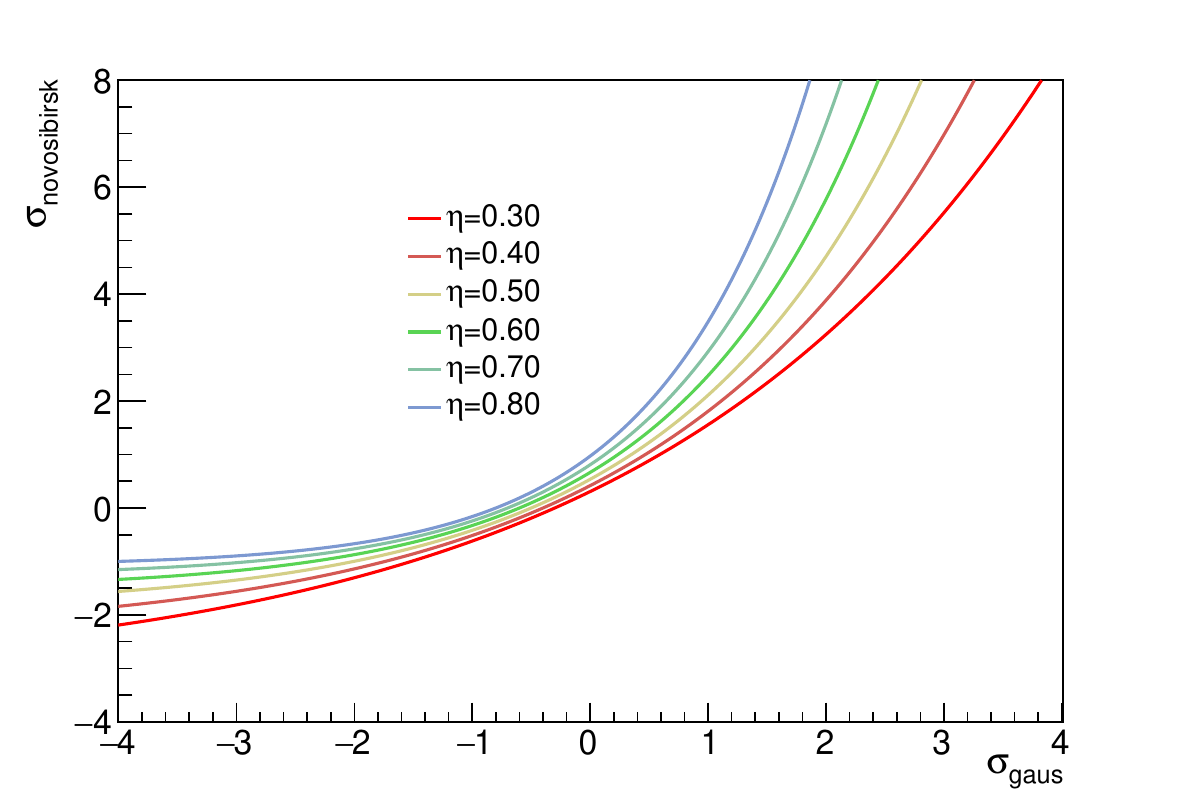}
    \caption{Inverse transformations $t^{-1}$ for different tail parameters. The $x$-axis gives the distance to the mean in units of $\sigma$ in the Gaussian variable which is mapped onto the distance to the MPV in units of $\sigma$ on the physical energy variable. With this function the fit internal variable $E'$ is translated back to physical energy values, e.g. the fit result, when the fit terminates.}
    \label{fig:inversetrafo}
\end{figure}

\noindent
The inverse transformation functions $t^{-1}(E')$ which are used to calculate the physical energy values $E$ from the transformed variable $E'$ are shown in Fig.~\ref{fig:inversetrafo}. It can be seen that a larger tail parameter leads to a larger curvature of the inverse transformation function. Note that the extrapolation  takes place outside of the shown region.
When integrating the variable transformation into the kinematic fit, the equations given in Section~(\ref{ch:theory}) must be adapted accordingly. In Equations~(\ref{eq:chi2}) and (\ref{eq:condition}), the energy variable $E'$ rather than $E$ needs to be used, and the covariance matrix now contains the Gaussian errors and  correlations of $E'$. \\

\noindent
After every iteration, the results for the energy, which serve as start values for the next iteration, need to be transformed back to the physical energy.
\begin{equation}
    E_s=t^{-1}(E'_s)
\end{equation}
\noindent
The same is done after the last iteration to obtain the final fit result. When performing the linearization (Eq.~\ref{eq:derivatives}), the expansion around the starting value has to be done using the transformed variable $E'$, and the derivative of the inverse transformation $t^{-1'}(E')$ has to be included accordingly. Eq.~\ref{eq:derivatives_new} shows the expansion around the start value. Note that the constraints are functions of the physical energy and not the transformed variable. 
\begin{equation}
\label{eq:derivatives_new} 
\begin{aligned}
    F_{iE'} =&\left . \frac{\partial f_i}{\partial E}\right|_{\vec{x}=\vec{x}_s} \cdot \left . \frac{\partial E}{\partial E'}\right|_{E'=E'_s} \\
    =& \left . \frac{\partial f_i}{\partial E}\right|_{\vec{x}=\vec{x}_s} \cdot t^{-1'}{}(E') 
\end{aligned}
\end{equation}
The vector $\vec{d}$ is still calculated as in Eq~(\ref{eq:derivatives_updated}). Here, $\vec{x_s}$ are the start values in the physical variable. It is important to note  that the transformation applied to the energy of a given photon is the same for every iteration and is always based on the likelihood derived from the measured value $E_0$. The expansion, however, must be performed for every intermediate fit result, and the functions $t^{-1}(E')$ and $t^{-1'}(E')$ must be known at every reasonable value for $E'$.

\subsection{Likelihood Spectrum}
\label{sec:likelihood}

\noindent
The transformation is defined using a likelihood distribution which becomes Gaussian in the transformed variable. Therefore, the shape of the likelihood distributions for all possible photons needs to be determined, which is done based on simulations.  \\
\begin{figure}[h]
    \centering
    \includegraphics[width=\columnwidth]{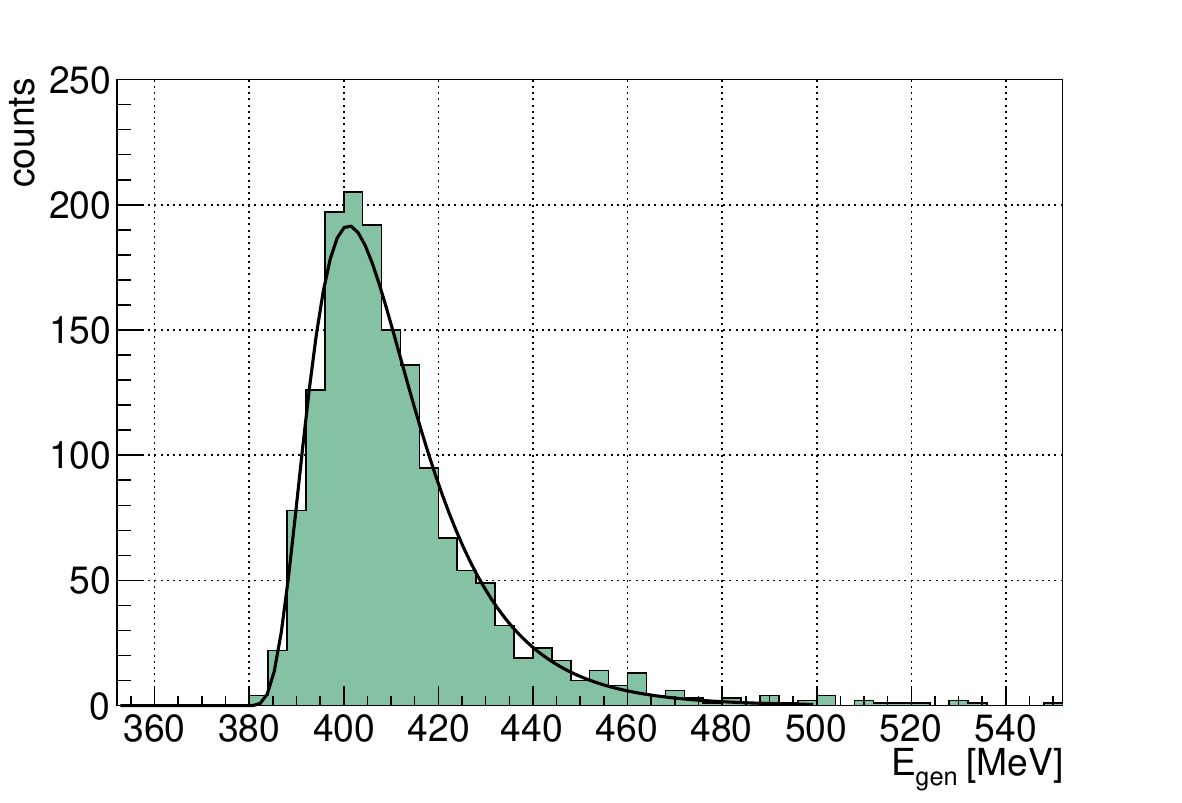}
    \caption{Likelihood distribution of the generated (true) photon energy $E_{\text{gen}}$ for photons measured with energies of $\SI{400}{\mega\electronvolt}\pm \SI{1}{\mega\electronvolt}$. This spectrum can also be fitted using the  Novosibirsk function from Eq.~(\ref{eq:Novosibirsk}). The fit is shown by the black line.}
    \label{fig:energylikelihood}
\end{figure}

\noindent
The likelihood spectrum of a specific measured energy $E_0$ (see Fig.~\ref{fig:energylikelihood}) is fitted with a Novosibirsk function, and its width $\sigma$ and tail $\eta$ parameters are determined. These spectra are obtained by simulating a continuum of photons with different energies over the whole solid angle covered by the detector, and then plotting, for a particular reconstructed angle, which of these generated energies result in a certain reconstructed energy $E_0$. As a test case, the Crystal Barrel Calorimeter of the CBELSA/TAPS experiment (see section~\ref{ch:CBELSA}) has been used. \\

\noindent
Note that in the case shown here, the MPV of the likelihood distribution lies (within the uncertainty) at the measured energy, meaning that when measuring a value of $\SI{400}{\mega\electronvolt}$, the most likely generated (true) value is also $\SI{400}{\mega\electronvolt}$. This follows from the fact that, in this example, the calibration is chosen such that the MPV of the PDF is located at the true value. In other words, when measuring a photon with an energy of $\SI{400}{\mega\electronvolt}$, the most likely measured value will also be $\SI{400}{\mega\electronvolt}$, which can be seen in Fig.~\ref{fig:energypdf}. We call this MPV calibration.\\

\noindent
If this is not the case for the data on which the transformation method is used, one must additionally determine the offset of the MPV position of the likelihood relative to the measured value. This must be included in the transformation function to shift the position of the Novosibirsk relative to the measured value $E_0$.

\subsection{MPV vs. Median}
\begin{figure}[h]
    \centering
    \includegraphics[width=1\columnwidth]{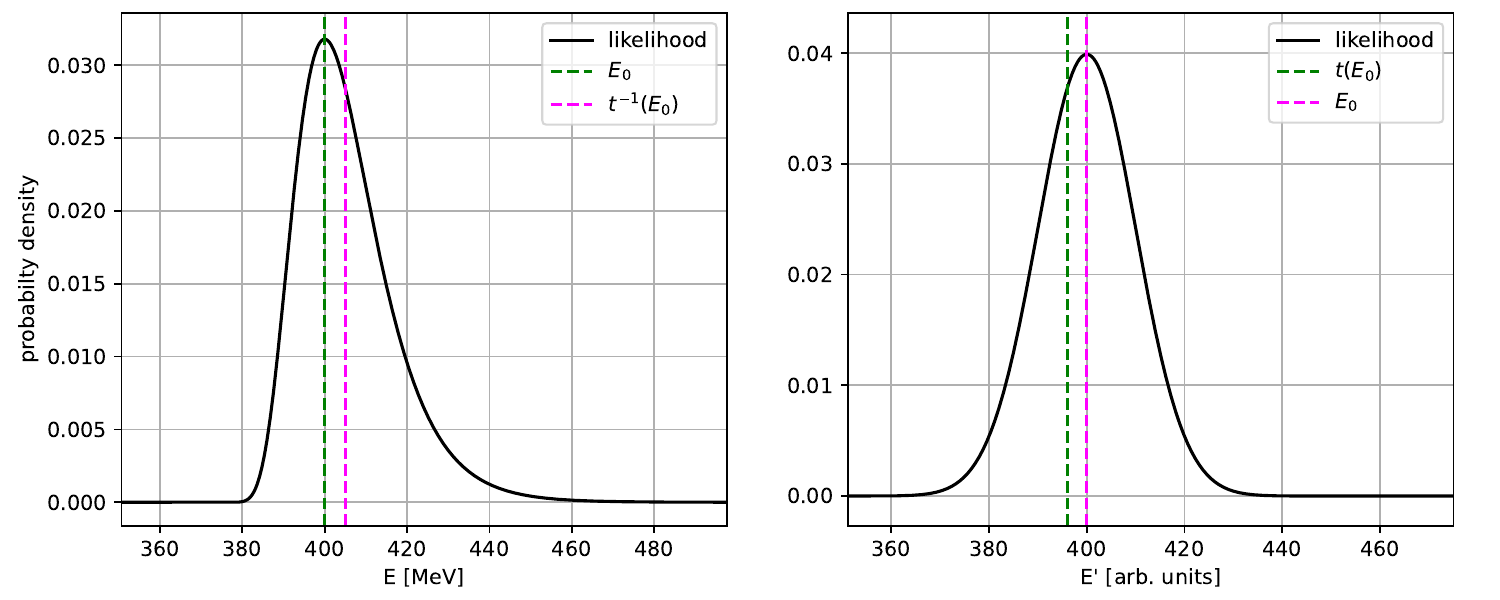}
    \caption{Comparison of a likelihood distribution before and after transformation. 
    Left: 
    the likelihood distribution on the physical energy $E$, right: 
    on the transformed energy variable $E'$. 
    Points that are mapped onto each other by the transformation $t(E)$ are indicated by a similarly colored line. As can be seen, the MPV of the likelihood on the left (green) is not mapped onto the center of the Gaussian. Instead the center of the Gaussian in $E'$ (magenta) corresponds to the median of the Novosibirsk distribution in the physical energy variable $E$.}
\label{fig:trafo_points}
\end{figure}

\noindent
When examining the transformation function, one finds that the MPV of the energy distribution is not mapped onto the most probable value of the corresponding Gaussian. This is an inherent feature of the approach since the transformation has to be continuous, and the median of the energy distribution must necessarily be mapped onto the mean of the Gaussian, which is also its median and MPV.
The situation illustrated  in Fig.~\ref{fig:trafo_points}. Here, a likelihood distribution and its transformation are shown for a measured energy of $\SI{400}{\mega\electronvolt}$. The points which are mapped onto each other are indicated by vertical lines of the same color.   

\quad\\
In cases where the MPV of the likelihood and the measured value align (MPV calibration), the physical energy value for which $\chi^{2}=0$ is not $E_{0}$, but rather the energy value that transforms to the MPV of the Gaussian variable. As  can be seen from the definition Eq.~(\ref{eq:inverse_trafo}) and Fig.~\ref{fig:trafo_points}, the transformation is constructed in such a way that the MPV of the Gaussian variable is always equal to the non-transformed measured energy value $E_0$\footnote{ This was a specific choice. Strictly speaking, there is no need to have the Gaussian in the $E^\prime$ variable also centred around $E_0$.}. Transforming this value back to the physical energy variable always gives a larger value than the original measured energy:
\begin{equation}
    \label{eq:MPV_median}
    E_{\chi^2=0}=t^{-1}(E'=E_0)>E_0 \, .
\end{equation}
So the consequence of using MPV calibrated energies is that, on average, the fit increases the energy of the measured photons based on the parameters $\sigma$ and $\eta$. A fit result with a $\chi^2$ of zero will return an energy larger than the measured value. The reason for this is that statistically, it is more likely that the measured energy is lower than the true energy.

\qquad\\
Here we have to emphasize that shifting the energies upwards relative to an MPV calibration is not unusual. Generally, electromagnetic calorimeters used in hadron physics are calibrated on the MPV of the $\pi^0$ mass peak, which leads to a calibration where the MPV of the reconstructed photon energy distribution lies higher than the true value. The differences between the MPV calibration on the level of the individual photons and a more common calibration on the mass of known particles, as well as how to obtain such an MPV calibration from data, are discussed further in the following section.

\subsection{Obtaining an MPV-calibration of  the EMC on real data}
\noindent
As established in Section~\ref{sec:likelihood}, MPV calibration refers to a calibration of the calorimeter where the MPV of the measured energy distribution for a given true energy aligns with the true energy. This can be seen in Fig.~\ref{fig:energypdf}, where the peak of the distribution lies at $\SI{400}{\mega\electronvolt}$, which was the energy of the generated photons.\\ 
Calorimeters in hadron physics experiments are generally calibrated on the mass of the $\pi^0$ such that the observed position of the $\pi^0$-mass peak (MPV) corresponds to the known mass of the meson. On the level of the energies of the individual photons, this causes the MPV of the measured photon energy distribution to be larger than the true photon energy. This is due to the discussed low energy tails of the measured photon energy distributions; the entire measured energy distribution must be shifted upwards to compensate for the prominent low energy tail. 
In the given example with a $\SI{400}{\mega\electronvolt}$ generated photon, the MPV of the PDF would then sit at $\approx\SI{410}{\mega\electronvolt}$. \\

\noindent
However, even when calibrating on meson masses, a calibration where the MPV of the measured photon energy distribution corresponds to the true energy value can be obtained. 
To do so, simulations of reactions including $\pi^0$s are performed using MPV-calibrated photon energies in the calorimeter. From those simulated data sets, the invariant mass $m(\gamma\gamma)$ can be reconstructed, which will show the $\pi^0$ peak significantly lower than the generated $\pi^0$-mass. 
Now an energy correction ($\rm ecf_\gamma^{\pi^0}$) can be applied, which increases the energy of the individual photons ($E_{\text{corr}}=E_{\text{MPV}}\cdot \rm ecf_\gamma^{\pi^0}$) such that the $\pi^0$ now lies at its correct mass. One simple approximate way of doing this correction is by shifting the photon energy upwards by a percentage of its uncertainty $E_{\text{corr}}=E_{\text{MPV}}(1+\alpha\sigma)$, with some empirical factor $\alpha$ that is tuned such that the $\pi^0$ peak lies at the desired position. 
Calibrating the real data on the $\pi^0$ mass while applying the energy correction factor $\rm ecf_\gamma^{\pi^0}$, determined from simulations, allows one to later obtain $E_{\text{MPV}}$ for the data by $E^{data}_{\text{MPV}} = E_\gamma^{data}/\rm ecf_\gamma^{\pi^0}$. 
\section{Application of the new Method}
\label{ch:Test}
\noindent
The new approach to kinematic fitting has been tested using simulated data of the CBELSA/TAPS experiment. However, before discussing the results of the test, we present a short overview of 
the CBELSA/TAPS experiment. 

\subsection{The CBELSA/TAPS Experiment}
\label{ch:CBELSA}
\noindent
The CBELSA/TAPS experiment is based at the electron accelerator ELSA in Bonn. It is a photoproduction experiment utilizing polarized beams and targets to study the spectrum and the properties of baryon resonances in the light u- and d-quark sector. 
The ability of the experiment to also perform measurements with polarized beam and target has been crucial in significantly improving our understanding of $N^*$- and $\Delta^*$-resonances and their properties. 
A detailed description of the experiment can be found in \cite{Gottschall21}. The experiment prominently features two electromagnetic calorimeters: the Crystal Barrel (CB) Calorimeter \cite{AKER1992} with 1320 CsI(Tl) crystals covering the entire $\phi$-angular range, as well as polar angles from $(12-156)^{\circ}$, and a wall of 216 BaF$_2$ crystals referred to as the TAPS spectrometer \cite{Novotny}, which covers the forward angular range down to $1^{\circ}$.\\

\noindent
Charged particles are identified by a three layer scintillating fiber detector surrounding the target~\cite{SUFT2005416} or by scintillators in front of the crystals~\cite{wendel, Novotny}. The energy of the incoming photons is determined using a tagging system~\cite{fornet-ponse}. 
The experiment is especially well suited for measuring photons from neutral meson decays, such as the $\pi^0$ or $\eta$, and therefore provides an ideal test case for the developed method.

\subsection{Determination of the Likelihood Parameters for the CB-calorimeter}
\label{ch:determination_parameters}
\noindent
To perform kinematic fitting using  the new transformation method, the width $\sigma$ and tail $\eta$ parameters of the likelihood distributions must be determined as a function of the measured energies and directions in the detector. The general method for the extraction of these parameters is outlined in section \ref{sec:likelihood}. This involves simulating individual photons at different energies and directions in the detector. This was done using a custom event generator, with the final tracking of the particles through the detectors performed in GEANT3.\\

\noindent
Because the Crystal Barrel detector is highly $\phi$--symmetric it can safely be assumed that the parameters only depend on the measured polar angle $\theta$ and the energy. 
Thus, the parameter values $\sigma$ and $\eta$ are determined as functions of energy in $1^\circ$ intervals in polar angle $\theta$. To do this, spectra at different energies and polar angles are first fitted by a Novosibirsk function. After that, for every polar angle interval, the energy dependence is fitted with an energy parametrization.\\

\begin{figure}[h]
    \centering
\includegraphics[width=\columnwidth]{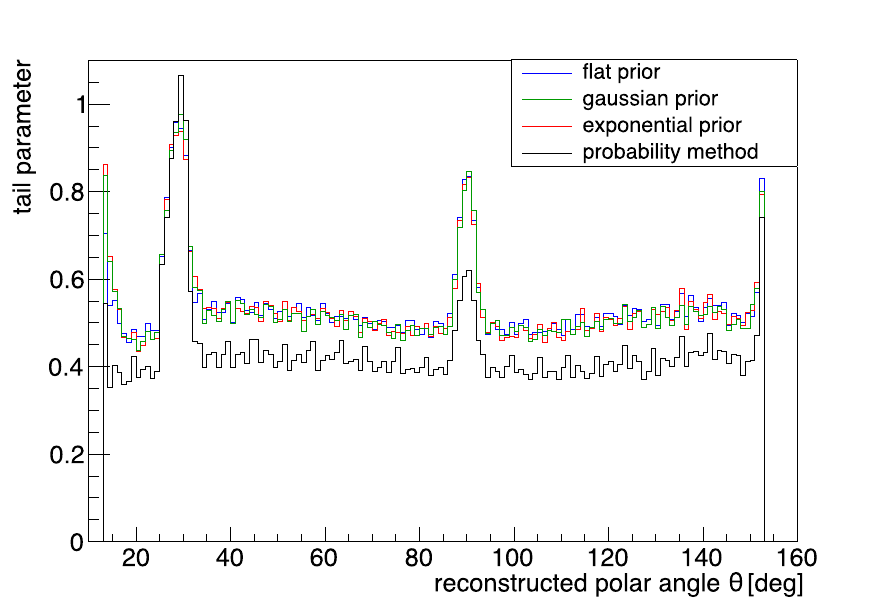}
    \caption{Extracted tail parameters at different polar angles, using different methods and prior energy distributions. The results of fits to the likelihood distributions using generated events with different energy distributions are shown in blue, green and red.  
    The black distribution shows the absolute values of the tail parameters as determined from a fit to the PDFs. See text for details.}
    \label{fig:tails}
\end{figure}

\noindent
For the widths $\sigma$, the observed energy dependence corresponds to the usual form for calorimeters, where the different energy dependence terms are added quadratically: $\left( \sigma/E \right)^{2} = \left(A/\sqrt{E}\right)^{2} + \left( B/E \right)^{2} + \left(C\right)^{2}$. The tail parameter does not show a strong energy dependence and was thus parametrized as constant in the fits for the energy dependence, but it does strongly depend on the polar angle $\theta$. The tail parameters become large in areas with more insensitive material or at the detector edges due to shower leakage. This can be seen at the edges and around $30^\circ$ and $90^\circ$ in Fig.~\ref{fig:tails}.\\

\noindent
One caveat to consider is that the likelihood distribution, in contrast to the PDF, generally depends on the true energy distribution of the events. This has been studied by investigating the influence of different prior photon energy distributions on the likelihood spectra. For the first extracted parameter set, the prior photon distribution has been taken to be flat in energy and solid angle ($2\pi\, d\cos{\theta}$). Even though this is not realistic, it represents a maximum entropy scenario where a minimum of input information is known. In addition, scenarios where the prior fell off exponentially in energy, or was distributed according to a broad Gaussian were tested. The choice of prior has not led to any noticeable systematic differences. Fig.~\ref{fig:tails} shows that the tail parameter sets determined from fits to the likelihood spectra with different priors  are very similar, which also holds for the extracted widths $\sigma$.
\quad\\

\noindent
For comparison, the tail parameters extracted from the PDFs (see Fig.~\ref{fig:energypdf}) are also shown in Fig.~\ref{fig:tails}. Since the tails of the likelihood distributions and the PDFs point in the opposite direction, $|\eta|$ is plotted. There are $10\%$ to $25\%$ differences between the values obtained from the two methods. This is in line with expectations (see section~\ref{ch:statistics});  the likelihood distribution is not obtained by merely flipping the PDF. 
However, the extracted width parameters $\sigma$ are more similar, with differences on the order of a few percent.

\subsection{Test of the new kinematic fitting approach}
\noindent
In the following, Monte Carlo simulated data from the CBELSA/TAPS experiment was used as a test case for the new approach to kinematic fitting. Here, the same simulation framework as in section~\ref{ch:determination_parameters} was utilized, with the custom event generator having been used to generate complete reactions instead of individual photons. For the tests performed in this section, only photons reconstructed with the Crystal Barrel calorimeter were considered. \\

\noindent
Simulated data is crucial to judge the performance of the fit as it is known if the fitted events belong to the correct hypothesis. It is only in this case that a flat confidence level with pulls that follow a normal distribution is expected. For real data, there will always be some background contamination, which makes the results more difficult to judge. Additionally, simulated data sets of different channels can be used to perform  studies on signal-to-background ratios in a controlled way.\\

\noindent
Various reactions with final states consisting of neutral pseudoscalar mesons ($\pi^0$, $\eta$) decaying into photons have been simulated. For final states with multiple mesons, the kinematic fit plays an especially important role since it   both significantly increases the resolution and reduces the combinatorial ambiguities in the final state.\\

\noindent
To test the new method, fits have been performed using the existing standard kinematic fitting procedure as well as the new one with the transformation of the energy variable. In the following, we refer to the standard kinematic fitting procedure \emph{KineFit} and the new method \emph{KineFit-Trafo} to indicate that it is based on the discussed energy transformation.\\

\noindent
Firstly, events of the reaction $\gamma \mathrm{p} \to \mathrm{p}3\pi^0\to\mathrm{p}6\gamma$ have been simulated, where the generated events follow a phase space distribution. On the simulated data, a four-constraint fit (4C-fit) has been performed. The initial number of constraint equations is seven, which come from the four-momentum conservation of the overall reaction and the masses of the three pions but, as the final state proton is treated as missing, the number of effective constraints is reduced by the three free parameters introduced by the momentum of the proton \footnote{The mass of the missing proton was fixed to the nominal value}.
Since it is a priori not known which of the six photons stems from which $\pi^0$ decay, all 15 possible combinations are fitted, and the one with the lowest $\chi^2$ value is taken as the correct combination.  
For the test, the exact energy of the initial beam photon is used, as the new method only relates to the measurement of final state photons. \\

\noindent
To judge the impact of applying the transformation, the confidence level as well as the pull of the energy are investigated, and the distributions before and after applying the transformation are compared. For these tests, a sufficiently high number of effective conditions is desirable to avoid a high correlation of the different pull distributions, which would make it more difficult to evaluate the effectiveness of the method. In the case of a less constrained fit, for example, an asymmetry in the energy distribution would also lead to an asymmetry in the polar angle pull, and vice versa.\\

\begin{figure}[h]
    \centering
    \includegraphics[width=\columnwidth]{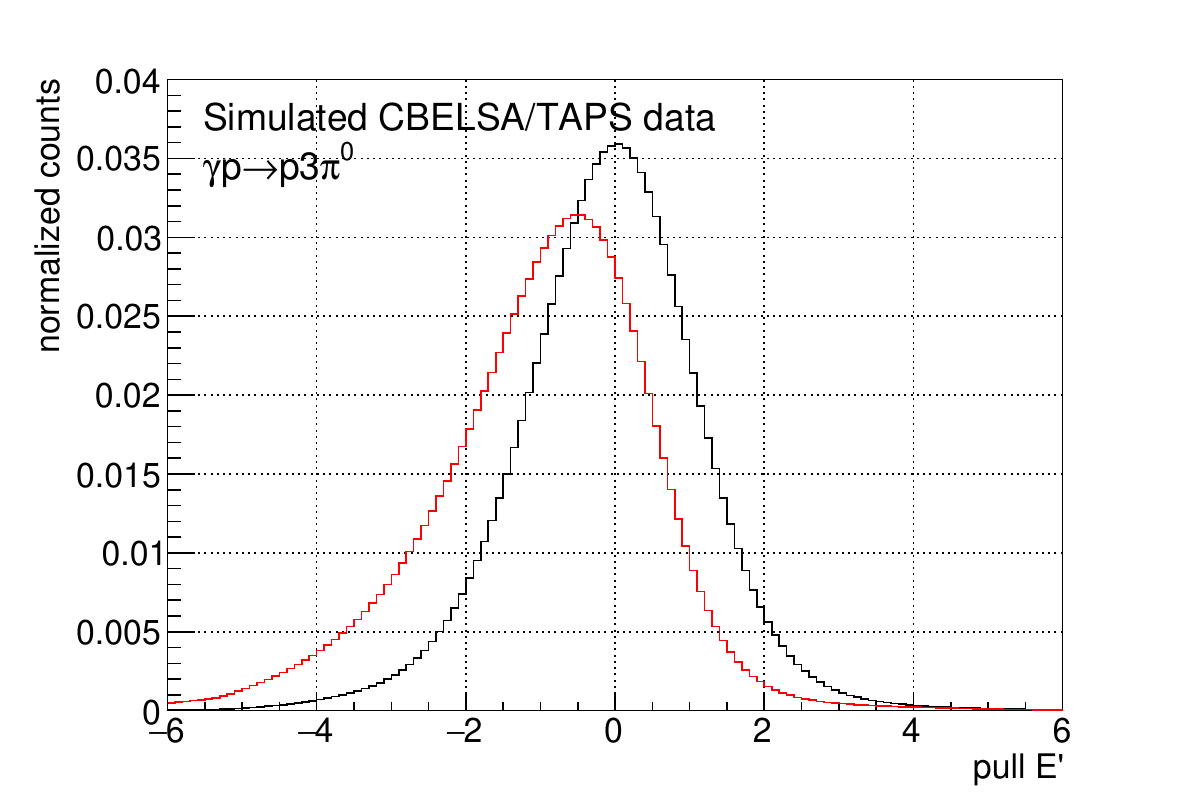}
    \caption{Pull distributions of the (transformed) energy variable for Monte Carlo simulated events of the reaction $\gamma\mathrm{p}\to\mathrm{p}3\pi^0$ fitted with the hypothesis $\gamma\mathrm{p}\to\mathrm{p}3\pi^0$. The events from the fit with energy transformation are shown in black and the results of the fit without transformation in red.}
    \label{fig:pulls3pi0}
\end{figure}

\noindent
In Fig.~\ref{fig:pulls3pi0} the photon-energy pull distributions resulting from a kinematic fit of the simulated data using the hypothesis $\gamma p \to p_{\text{miss}}\, 3\pi^0 \to p_{\text{miss}}\, 6\gamma$ are shown. The results of the KineFit-Trafo fit using the transformed energy variable $E'$ (black) are compared to the results of the standard KineFit (red). A clear improvement is observed;  the shape of the KineFit-Trafo pull is less asymmetric and generally much closer to that of a normal distribution than the pull from the KineFit case. Of note is that the MPV of the pull for the fit with energy transformation is much closer to zero than previously. A very similar improvement is also observed for other final states.\\

\begin{figure}[h]
    \centering
    \includegraphics[width=\columnwidth]{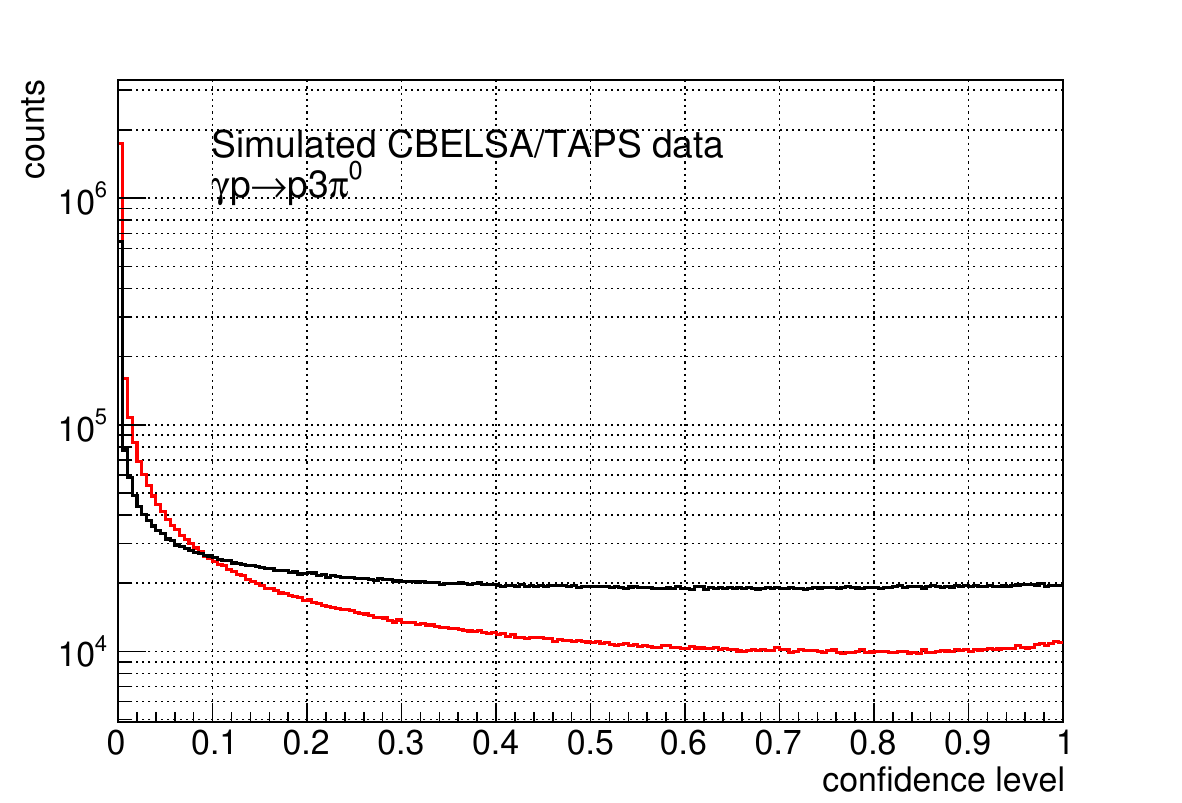}
    \caption{Confidence level for Monte Carlo simulated events of the reaction $\gamma\mathrm{p}\to\mathrm{p}3\pi^0$ fitted with the hypothesis $\gamma\mathrm{p}\to\mathrm{p_{miss}}3\pi^0$. The events from the fit with energy transformation are shown in black, the results of the fit without transformation in red.}
    \label{fig:cl_compare}
\end{figure}

\noindent
In addition to the pull distribution, the confidence level distribution confirms the improvement due to the energy transformation, as shown in Fig.~\ref{fig:cl_compare}. The region where the confidence level is flat becomes larger, and the number of events at very low confidence levels is significantly reduced by a factor of three in the lowest confidence level bin. Still, even in the transformed case, there remains an excess at small confidence levels; this is discussed further in section~\ref{sec:discussion}. \\

\noindent
The impact of the transformation on the actual fit result also becomes clearly visible in the invariant mass distributions of intermediate particles that are not constrained by the kinematic fit. Here, kinematic fitting is especially important as the mass resolution of these particles can be {significantly improved. One example is the reaction $\gamma \mathrm{p} \to \mathrm{p}\, \eta \to \mathrm{p}\, 3\pi^0\to\mathrm{p}\, 6\gamma$, where the three neutral pions stem from the decay of an $\eta$-meson. This reaction has been simulated and then fitted with the masses of the three $\pi^{0}$s constrained. Fig.~\ref{fig:etacompare} shows the invariant mass peak of the $\eta$ meson after applying the kinematic fit. Compared to the KineFit method, the peak for the KineFit-Trafo fit is more symmetric, centered at the nominal $\eta$ mass, and has a larger yield. In the Kinefit-result, a significant number of events in the peak lie in a tail that extends to lower masses. The advantage of the new method becomes clearly visible. \\

\begin{figure}[h]
    \centering
    \includegraphics[width=\columnwidth]{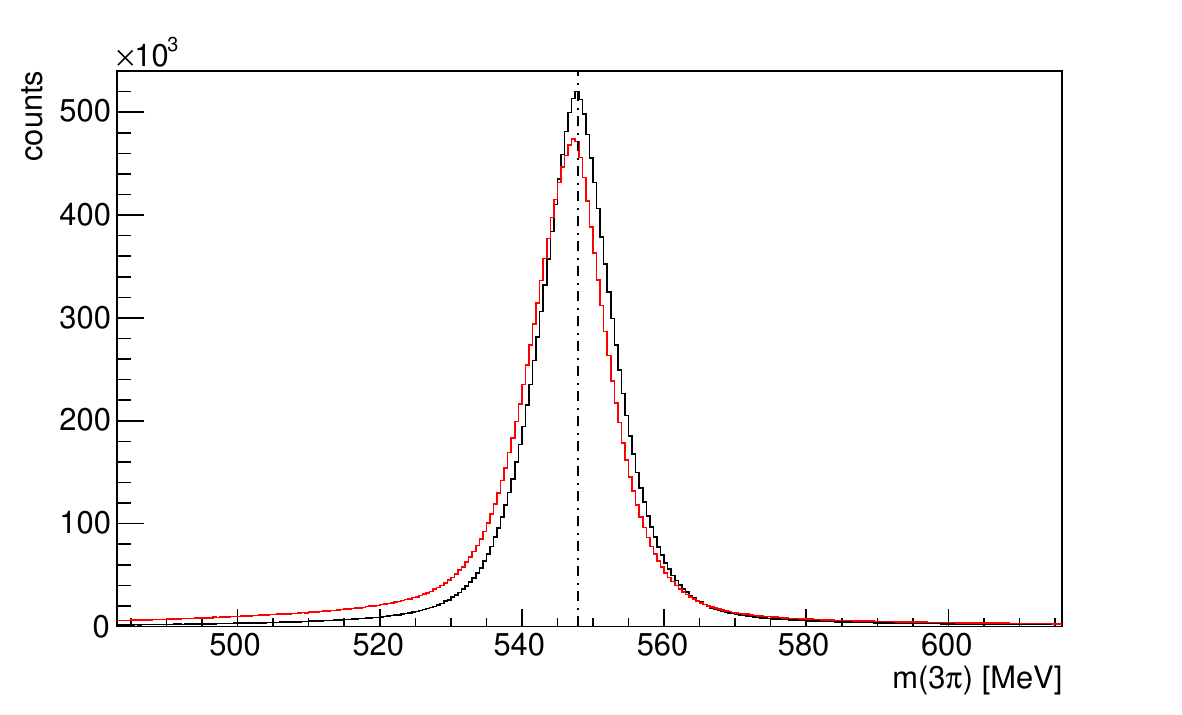}
    \caption{Invariant mass of the $3\pi^0$ system for Monte Carlo simulated events of the reaction $\gamma\mathrm{p}\to\mathrm{p}\eta\to\mathrm{p}3\pi^0\to\mathrm{p}6\gamma$ fitted with the hypothesis $\gamma\mathrm{p}\to\mathrm{p_{\text{miss}}}3\pi^0\to\mathrm{p_{\text{miss}}}6\gamma$. The events from the fit with energy transformation are shown in black and the results of the fit without transformation in red. The dotted-dashed line shows the nominal mass of the $\eta$~\cite{PDG}.}
    \label{fig:etacompare}
\end{figure}

\noindent
In the analysis of experimental data, the confidence level is often used as a cut criterion. Here, the goal is to choose a cut that retains the most signal while also achieving the strongest background rejection, thus optimizing the signal-to-background ratio.
To evaluate how the energy transformation impacts background rejection, the $\mathrm{p}\pi^0 \pi^0 \eta$ channel has been investigated. In experimental data, this channel contains a significant background from the $\mathrm{p}\,3\pi^0$ channel. 
For this reaction, two fits can be performed. The first with the hypothesis of the desired signal and the second with the hypothesis of the background channel. Cuts can then be made on both hypotheses. For the signal hypothesis, the confidence level should lie above a certain threshold $p_{\text{sig}}$ to keep the good signal events while rejecting those of poor quality. The background-channel hypothesis should be unlikely for the signal events so that the confidence level of this fit should be smaller than some value $p_{\text{bg}}$. By setting $p_{\text{bg}}$ to $0.01$ and varying the cut on the confidence level of the fit to the signal hypothesis, the percentage of retained signal events for different remaining background fractions has been determined (see Fig.~\ref{fig:signaltonoise}). The remaining signal fraction for the KineFit-Trafo fit is significantly larger for any given background level. The same general behavior is also observed when setting $p_{\text{bg}}$ to different values. 
\begin{figure}[h!]
    \centering
    \includegraphics[width=\columnwidth]{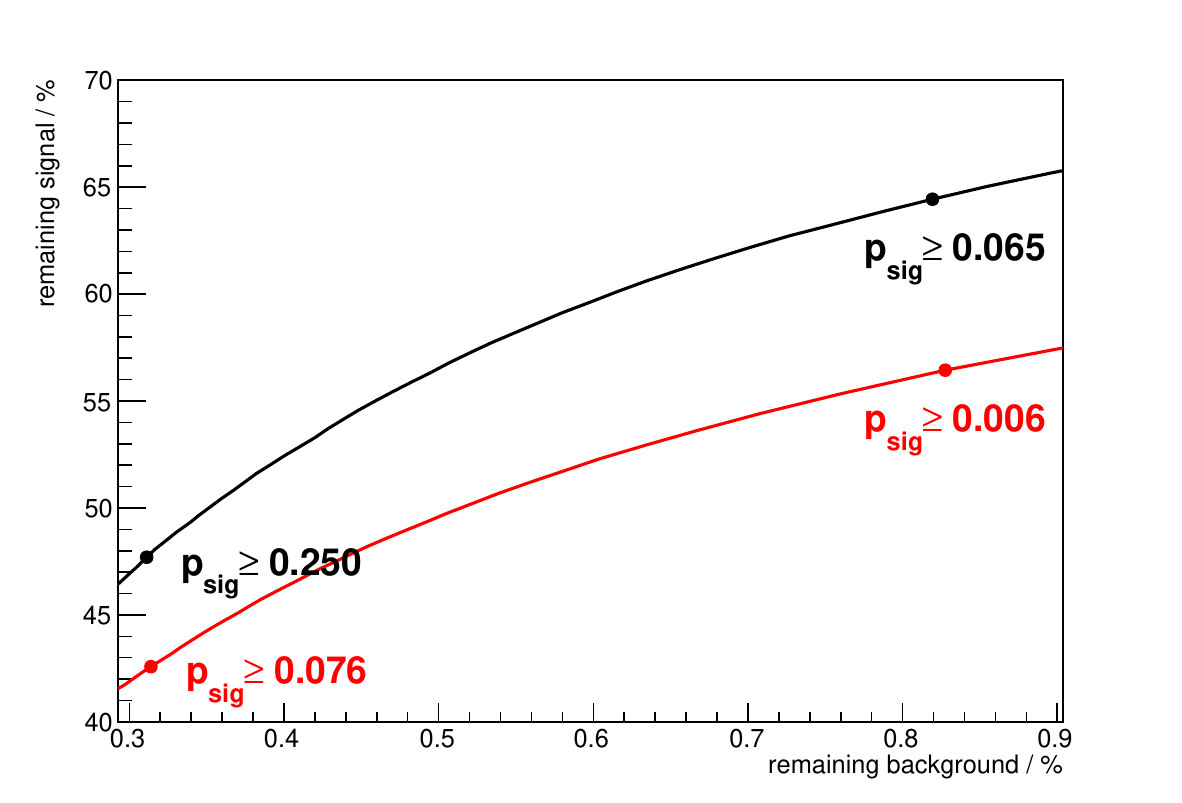}
    \caption{Remaining $p\pi^0 \pi^0 \eta$ signal vs remaining $p\pi^0\pi^0\pi^0$ background percentage after applying different cuts to the confidence level of the fit to the signal hypothesis.  The anti-cut on the fitted background channel was fixed to $p_{\text{bg}}\leq 0.01$ The black curve shows the result for the fit with the energy transformation while the red curve shows the result without. The circular markers indicate points computed with a given cut on $ p_{\text{sig}}$.}
    \label{fig:signaltonoise}
\end{figure}

\section{Summary and discussion}
\label{sec:discussion}
\noindent
As shown in the previous section, the energy transformation has a significant positive impact on statistical quantities such as the pull and the confidence level. In addition, the actual physical fit result obtained from the kinematic fit improves. \\

\noindent
The observed changes in the pull distributions clearly show that including the transformation improves the statistical description of the data, as the inherent assumption of Gaussian fit quantities is now much better fulfilled. This is also seen in the confidence level, which now remains flat over a wider region and has significantly fewer signal events at very low confidence levels. \\

\noindent
It has also been found that the observed masses of intermediate particles, which are not constrained, are closer to their nominal values\cite{PDG} with peaks that display less asymmetry. This improves signal yields when cutting on invariant mass peaks of intermediate particles, as fewer events lie in the tail of the distribution (see Fig.~\ref{fig:etacompare}).\\

\noindent
Using the simulated data for the photoproduction of $3\pi^0$ and $2\pi^0\eta$ off the proton, it has been shown that the KineFit-Trafo method improves the signal to background ratio relative to the KineFit method. 
This is due to the fact that the confidence level now more accurately describes the quality of an event, and fewer good signal events are observed at very low confidence levels. As most of the background events are found at very low confidence levels, this effectively improves the discriminating power of the fit. \\

\noindent
When using the KineFit-Trafo method, the confidence level of the fit result is still not perfectly flat. Some residual curvature is observed, and an excess of events at very low confidence levels remains. Four reasons for this have been identified:\\
\begin{itemize}
\item[(i)] The description of the energy distributions via the Novosibirsk function is only a model. It does not account for all effects which might negatively affect the reconstruction. Generally, there is a sizable number of events that are poorly reconstructed, even in comparison to the expected low energy tail. This happens especially in areas of the detector system with an increased amount of insensitive material or at the edges of the detector system where part of the electromagnetic shower is lost. Here, an excess of events at very low energies close to the reconstruction thresholds is observed. Rejecting these events is reasonable since such events cannot be reliably distinguished from background in the real data.
\item[(ii)] In addition, the extraction of the likelihood parameters has systematic errors since the fits to the likelihood distributions are only performed for a finite number of points in energy and direction. The functions used to describe the energy dependence only approximate the full dynamics of the parameters.
\item[(iii)] Furthermore, even in a perfect model, where all the input distributions are perfectly known, there is still a small excess at very low confidence levels. This is due to an inherent limitation of the kinematic fit method, as the expansion of the condition Eq.~(\ref{eq:derivatives}) is only a good approximation as long as the difference between the measured and true values of any given variable is small. Therefore, for the events with the largest deviations from the true value, the fit does not necessarily converge to the correct global minimum.
\item[(iv)] Finally, the simulations performed for  the  CBELSA/TAPS experiment use a $\SI{5}{\centi\meter}$ long hydrogen target, while in the reconstruction of the photon polar angle, it is assumed that the photon originated from the center of the target. This is addressed by choosing the errors of the polar angles accordingly. Nevertheless, this still leads to a distortion of the confidence level, as larger uncertainties on the polar angles do not correctly account for the fact that the polar angle errors resulting from the vertex displacement are highly correlated.
\end{itemize}

\noindent
The influence of the prior photon distribution on the energy errors $\sigma$ and tail parameters $\eta$ extracted from the likelihood distributions is found to be  small. Therefore, a single parameter set is determined by fitting the likelihood functions for the different ($E_\gamma, \theta)$-bins of the CBELSA/TAPS-calorimeter, and these parameters have been used for all reactions studied. \\

\noindent
Based on the tests performed, it has been shown that the transformation of the energy variable leads to a significant improvement in all statistical metrics and an improved quality of the final fit result.\\

\section{Conclusion}
\noindent
By applying a transformation to the energy variable in the kinematic fit, one can remedy the effect of highly asymmetric energy distributions on the fitting procedure. This leads to increased robustness of the statistical measures, such as the confidence level, as well as improvements in the overall fit results.   

\section*{Acknowledgements}
\noindent
We thank the CBELSA/TAPS collaboration for the opportunity to use their simulation and reconstruction software as the basis for the 
studies performed within this paper. The authors thank C.~A.~Meyer for carefully reading the manuscript, many helpful comments and fruitful discussions. In addition, we thank Ph.~Mahlberg, and C.~M.~Frenkel for their contributions.  
This work was supported by the Deutsche Forschungsgemeinschaft (DFG, German Research Foundation) under Germany's Excellence Strategy - Cluster of Excellence "Color meets Flavor", EXC 3107 - Project-ID 533766364 
This work is part of the Ph.D. thesis of N. Kolanus. 

\section*{Appendix: Decomposition using orthogonal transformations}
\label{ch:Appendix}
\renewcommand{\theequation}{A.\arabic{equation}}
\setcounter{equation}{0}
\noindent
The minimization algorithm based on orthogonal transformations was first described by Lawson and Hanson~\cite{LAWSON74}, while the description provided here follows the one given in \cite{brandtDataAnalysisStatistical2014}. In this method, the minimization problem is solved by first performing a so-called LQ decomposition of the $(m\times n)$ matrix $F$:
\begin{equation}
    \label{eq:qr-decomp}
    F=LQ^T \, .
\end{equation}
In doing this separation, one can always find an orthogonal $(n\times n)$ matrix $Q$ such that the $(m\times n)$ matrix $L$ can be split column-wise into a $(m\times m)$ lower triangular matrix $L_1$ and a $(m\times(n-m))$ zero matrix $L_2$:
\begin{equation}
    \label{eq:column-wise}
    L=(L_1,L_2)=(L_1,0) \, .
\end{equation}
This is useful, as on the transformed space 
\begin{equation}
    \vec{\delta}^{\prime}\, = \, Q^{T}\vec{\delta}\, ,
\end{equation} 
Eq.~\ref{eq:condition} from the main text, repeated here as Eq.~\ref{eq:conditiona}
\begin{equation}
\label{eq:conditiona}
F\vec{\delta} - \vec{d} \, = \, 0 \, ,
\end{equation}
can thus be rewritten
\begin{equation}
    LQ^{T} \vec{\delta} - \vec{d} \, = \, L\vec{\delta}^{\prime} - \vec{d} \, .
\end{equation}
we can now use Eq.~\ref{eq:column-wise} to write this as
\begin{equation}    
L\vec{\delta}^{\prime} - \vec{d} \, = \, \left( L_{1}\vec{\delta}^{\prime},L_{2}\vec{\delta}^{\prime}\right ) -\vec{d} 
\end{equation}
which now simplifies to:
\begin{equation}
    \label{eq:r-matrix}
    L\vec{\delta'}-\vec{d} \, = \, L_1\vec{\delta_1'}-\vec{d}\, ,
\end{equation}
and noting this is zero from Eq.~\ref{eq:conditiona}, we have
\begin{equation}
\label{eq:sperated_condition}
L_{1}\vec{\delta}_1^{\prime} - \vec{d} \, = \, 0 \, .
\end{equation}
We note that $\vec{\delta}_1'$ refers to the first $m$ components of the $n$-vector $\vec{\delta}'$. \\

\noindent
This decomposition can be nicely demonstrated using a low dimensional example. Given only two fit quantities $\delta_1$ and $\delta_2$, one can reasonably apply one constraint. For a linear constraint in two dimensions, all solutions lie on a line (i.e., in a 1D subspace), as shown in Fig.~(\ref{fig:q_rotation}). 
The constraint is then given by the $(1\times2)$ matrix $F$ and the scalar $d$.\\
\begin{equation}
    F\vec{\delta} - d \, = \, \left( F_{11} , F_{12} \right ) \, \left( \begin{matrix} \delta_{1} \\ \delta_{2} \end{matrix} \right ) - d \, = \, 0
\end{equation}
which can be expanded as
\begin{equation}
    F\vec{\delta} - d = F_{11}\delta_{1} + F_{12}\delta_{2} - d = 0 \, .
\end{equation}
The LQ decomposition of $F$ can now be formulated as the following problem. One wants to find a rotation Q so that, in the resulting coordinate system, one of the two axes lies parallel to the line given by the constraint. This means that this axis is now completely independent of the constraint, whereas on the other axis, the equation is trivial to solve. The consequence is that the matrix $L$, which is now the constraint matrix in the rotated space, has a zero column for every unconstrained axis. By choosing an appropriate basis, one can always sort these columns to the right side of the matrix. From this follows the column-wise decomposition of the matrix $L$ into the lower triangular matrix $L_1$ and the zero matrix $L_2$. In this example, the matrices $L_{1}$ and $L_{2}$ into which $L$ can be decomposed, and which we notate as $L=(L_{1},L_{2})$, are $(1\times1)$ matrices, as shown in Eq.~(\ref{eq:decomp_example}).
\begin{equation}
    \begin{aligned}
     0=&F\vec{\delta} - d = L\vec{\delta}^{\prime} -d\\ =&
    ( 
    \underbrace{L_{11}}_{L_{1}} \, , \,  \underbrace{0}_{L_{2}}    
    )
    \left( 
    \begin{matrix}
    \delta^{\prime}_{1}\\
    \delta^{\prime}_{2}
    \end{matrix}
    \right) -d =
    L_{1}\delta^{\prime}_{1} -d   
    \end{aligned}
    \label{eq:decomp_example}
\end{equation}

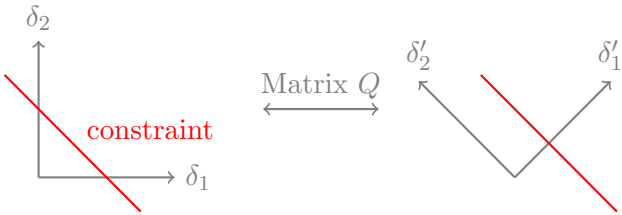
\begin{figure}[h]
    \centering
     \begin{tikzpicture}[scale=0.9]
        \draw[gray, thick,->] (0,0) -- (0,2) node[anchor=south]{$\delta_2$};
        \draw[gray, thick,->] (0,0) -- (2,0) node[anchor=west]{$\delta_1$};
        \draw[red, thick] (-.5, 1.5) -- (1.5,-0.5)  (0.55,0.7) node[anchor=west]{constraint};

        \draw[gray,thick,<->] (3.3,1) -- (5,1);
        \draw[gray] (4.15,1) node[anchor=south]{Matrix $Q$};

        \draw[gray, thick,->] (7,0) -- (5.59,1.41) node[anchor=south]{$\delta'_2$};
        \draw[gray, thick,->] (7,0) -- (8.41,1.41)  node[anchor=south]{$\delta'_1$};
        \draw[red, thick] (6.5, 1.5) -- (8.5,-0.5);

    \end{tikzpicture}

    \caption{Example of an LQ decomposition in 2D. The matrix Q rotates to a coordinate system where one of the coordinates is not constrained.}
    \label{fig:q_rotation}
\end{figure}
\noindent
The method \texttt{CompleteOrthogonalDecomposition()} from the \texttt{Eigen} library \cite{eigenweb} was used to implement the decomposition into the kinematic fit. \\

\noindent
In general, after such a decomposition has been performed, the equation $L_{1}\vec{\delta^{\prime}_{1}}-\vec{d}=0$ becomes trivial to solve and has a unique solution, as $L_{1}$ is a triangular matrix. At the same time, this transformation completely separates the unconstrained $n-m$ dimensional subspace $\vec{\delta^{\prime}_{2}}$ from the unique solution $\vec{\bm\delta^{\prime}_{1}}$ and the general solution of Eq.~(\ref{eq:conditiona}) $\vec{\bm\delta}$ can be written as:
\begin{equation}
    \label{eq:general_solution}
    \vec{\bm\delta} = Q_1\vec{\bm\delta^{\prime}_{1}} + Q_2\vec{\delta^{\prime}_{2}},
\end{equation}
where $\vec{\bm\delta^{\prime}_{1}}$ is the unique solution of Eq.~(\ref{eq:sperated_condition}) and $\vec{\delta^{\prime}_{2}}$ is an arbitrary vector in the unconstrained subspace. $Q_{1}$ and $Q_{2}$ result from the column-wise decomposition of $Q$ in the subspaces $1$ and $2$ as $Q=(Q_{1},Q_{2})$. 

\noindent
As only the elements in subspace $2$ are not fixed by the conditions of Eq.~(\ref{eq:conditiona}), the minimization must be performed only in this subspace of the transformed coordinate system. With this, the minimization problem can be written as:
\begin{equation}
\begin{aligned}
        \text{min}[\chi^2] =& \text{min}[(\hat{A}_1\vec{\bm\delta^{\prime}_{1}}+\hat{A}_2\vec{\delta^{\prime}_{2}})^{2}]\\ =& \text{min}[(\hat{A}_2\vec{\delta^{\prime}_{2}}-\vec{a})^{2}]
,
\end{aligned}
\label{mini}
\end{equation}
where $\vec{a}=-\hat{A}_1\vec{\bm\delta^{\prime}_{1}}$ is used to bring the problem into its generic form. From Eq.~(\ref{mini}) a unique solution $\vec{\bm\delta^{\prime}_{2}}$ is calculated. By inserting both $\vec{\bm\delta^{\prime}_{1}}$ and $\vec{\bm\delta^{\prime}_{2}}$ into Eq.~(\ref{eq:general_solution}), the unique solution for $\vec{\bm\delta}$ is obtained, and the minimization problem is solved.

\printbibliography

\end{document}